\documentclass[useAMS,usenatbib]{mn2e}
\usepackage{graphicx}

\usepackage{graphics,color}
\definecolor{dark}{gray}{0.5}
\definecolor{red}{rgb}{1,0,0}
\definecolor{green}{rgb}{0,1,0}
\definecolor{blue}{rgb}{0,0,1}
\title{Galactic Parameters from Whole Sky 2MASS Star Counts}
\author[Chang et al.]{C.K. Chang$^1$, C.M. Ko$^{1,2}$ and T.H. Peng$^1$
\thanks{E-mail: rex@astro.ncu.edu.tw (CKC)} \\
$^1$  Institute of Astronomy, National Central University, Taiwan \\
$^2$  Department of Physics and Center for Complex Systems, National
Central University, Taiwan}

\begin{document}

\date{}

\pagerange{\pageref{firstpage}--\pageref{lastpage}} \pubyear{2002}

\maketitle

\label{firstpage}

\begin{abstract}
The whole sky differential star counts (DSC) with 1 degree
resolution are retrieved from 2MASS online data service. Galaxy with
double exponential thin and thick disks and a single power law
luminosity function (LF) is used to interpret the 2MASS data. The
slope of the DSC appears roughly isotropic over the whole sky, the
average value is $\sim 0.32$, which corresponds to a power law index
$\sim 1.8$ of the LF. We find that the scale-length and scale-height
the thin disk are $\sim 3.0$ kpc and $\sim 245$ pc, and those of the
thick disk are $\sim 3.0$ kpc and $\sim 780$ pc. The ratio of the
thick disk to the thin disk is $\sim7$\%. The location of Sun above
the disk is $\sim 15$ pc. A comparison of the data and model and
their discrepancy are also provided.
\end{abstract}


\section{Introduction}
Star counts provides us a good way to investigate the structure of
the Milky Way. In 1980 Bahcall \& Soneira proposed a two component
model, disk and halo, to interpret Galactic structure (Bahcall \&
Soneira 1980). However, in 1983 Gilmore \& Reid identified the disk
could be further divided into two parts, thin and thick, for better
fitting (Gilmore \& Reid 1983). Since then, many works based on
different observations with limited or specific sky area have been
carried out (Karaali et al. 2004) and indicate that the thick disk
is an individual component. The scale-heights of the thin disk in
these works are similar, ranging from 240 kpc to 350 kpc. However,
the scale-height of the thick disk have a much larger range and
tends to have smaller value with larger local stellar number density
(Chen et al. 2001). The Two Micron All Survey, 2MASS (Skrutskie et
al. 2006), provides an opportunity to use large sky coverage star
counts to investigate the Galactic structure and more detail
structures were recognized. The warp found on 2MASS is the same as
that from gas and dust. Also, the scale-height of the thick disk
increases with the Galactocentric radius for $R>5$ kpc
(Cabrera-Lavers et al. 2007). However, the best model to describe
the major smooth components, thin and thick disks, of the Milky Way
has not been fixed yet.

In this contribution, we use the whole sky 2MASS point source
catalog (hereafter, 2MASS PSC) to derive the overall structural
parameters and the luminosity function of the Milky Way.

\section{2MASS Point Source Catalog}
Sources with SNR $\geq 5$ and detected in all J, H, Ks bands were
selected from 2MASS PSC. Its completeness rate corresponding
$10\sigma$ SNR limiting magnitude is 99\%. By Hierarchical
Triangular Mesh, the whole sky is divided into 32768 nodes with
$\sim 1$ degree resolution. The data was retrieved directly from
2MASS online data service via virtual observatory protocol. We only
use Ks band data in this work. Due to the small number of stars
brighter than 8 magnitude and incompleteness fainter than limiting
magnitude, we analyze data between 8 to 14 Ks magnitude only. The
differential star counts (hereafter DSC) is well approximated by a
single power law in this region (see Fig.1 left).

%----------------------------------------------------------------------
\begin{figure} %Fig.~1
 %\centerline{{\epsfxsize=4cm\epsffile{dataModel.eps}\epsfxsize=4cm\epsffile{dips.eps}}}
 \centerline{\includegraphics[width=40mm]{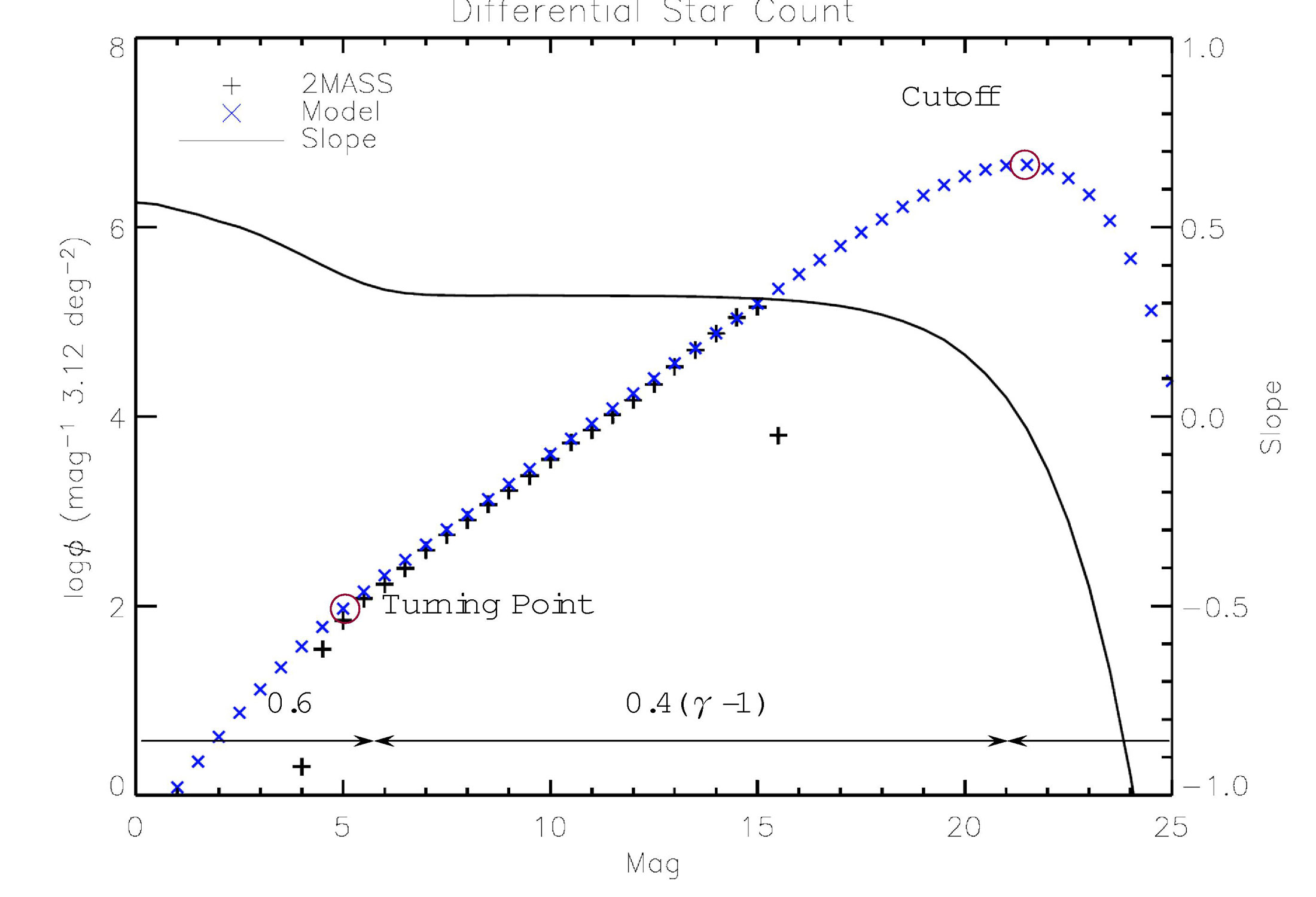}\includegraphics[width=40mm]{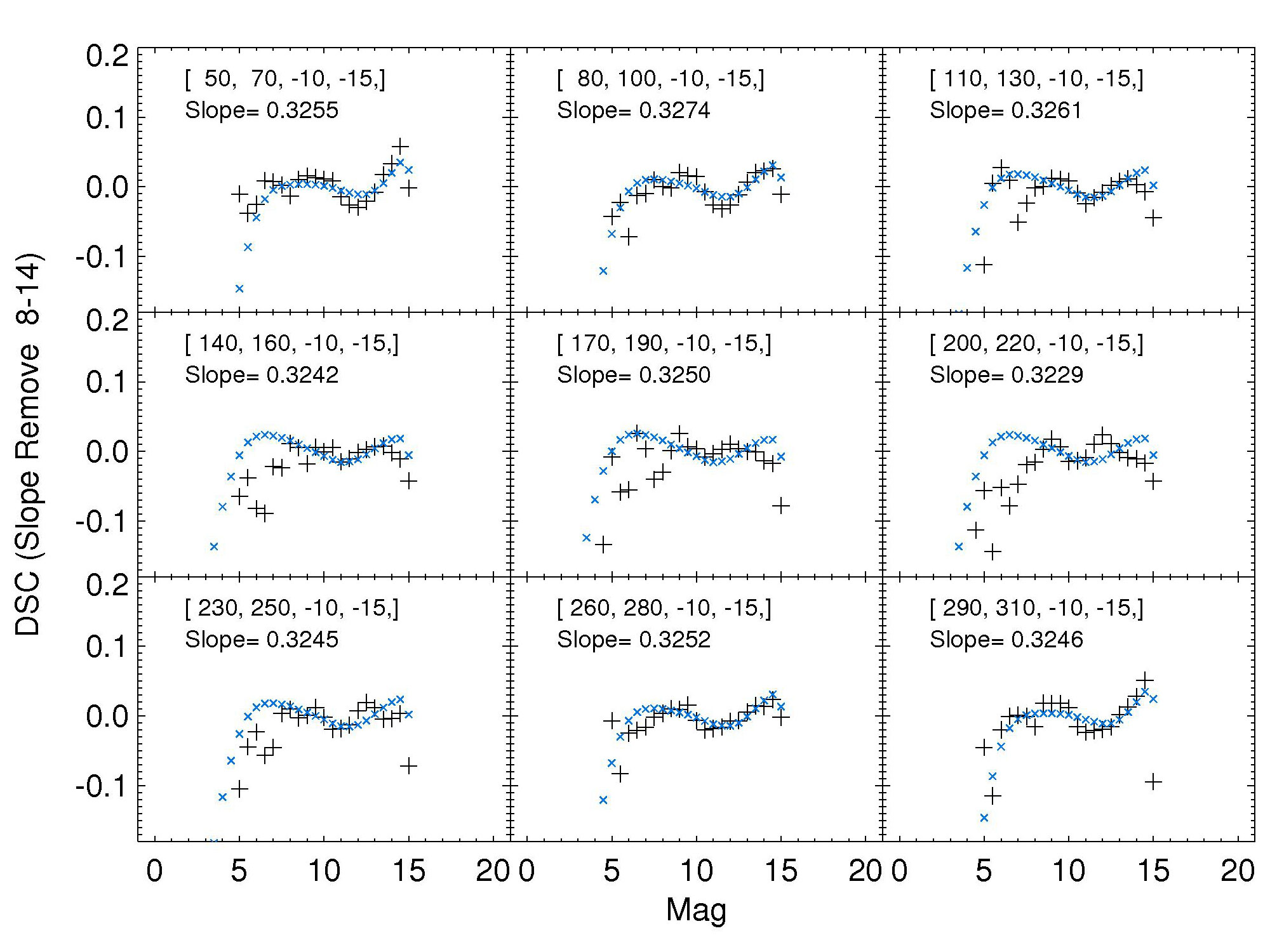}}
 \caption[]{Left:The Ks band DSC of 2MASS (black plus) and model (blue cross)
  at [l,b]=[270,10].
  Right:After removing the slope, DSC clearly shows a dip.
 The black plus is 2MASS data, and the blue cross is model.}
\end{figure}
%----------------------------------------------------------------------

\section{The model and its features}
Star count depends on the distribution of stars in space and
luminosity, i.e., stellar number density profile of the Galaxy and
local luminosity function. The density profile, $n(h_r,h_z)$, in our
model contains a thin disk and a thick disk. Both are exponentially
decayed along the Galactocentric radius, $R$, and the distance above
or below the Galactic plane, $Z$,
\begin{equation}
  n(r,z)=n_0\exp\left(-\frac{R-R_\odot}{h_r}\right)
  \exp\left(-\frac{|Z-Z_\odot|}{h_z}\right)\,,
\end{equation}
where $h_r$ and $h_z$ are the scale-length and scale-height of the
disk, $n_0$ is the local density in the solar neighborhood,
$R_\odot$ and $Z_\odot$ are the sun location. In our model the
luminosity function (hereafter LF), $\phi(L_1<L<L_2)\propto
L^{-\gamma}$, is a single power law, where $L_1$, $L_2$ are lower
and upper cutoffs.
%$\gamma$ is the power law index.
Moreover, we assume that the LF does not depend on location. The
synthetic DSC can be divided into 3 regions, bright, modest and
faint, with slopes 0.6, $0.4(\gamma-1)$ and a cutoff, respectively
(See Fig.1 left). Therefore, we can derive $\gamma$ from the slope
in the modest region. Moreover, the turning point and cut off can be
determined by $h_r$, $h_z$, $L_1$ and $L_2$. The New COBR/IRAS
extinction model (hereafter NCI model) was adapted for Galactic
latitude $|b|\geq 10^0$ area (Chen et al. 1999).

%----------------------------------------------------------------------
\begin{figure} %Fig.~2
 %\centerline{{\epsfxsize=4cm\epsffile{allskySlope.eps}}}
 \centerline{\includegraphics[width=80mm]{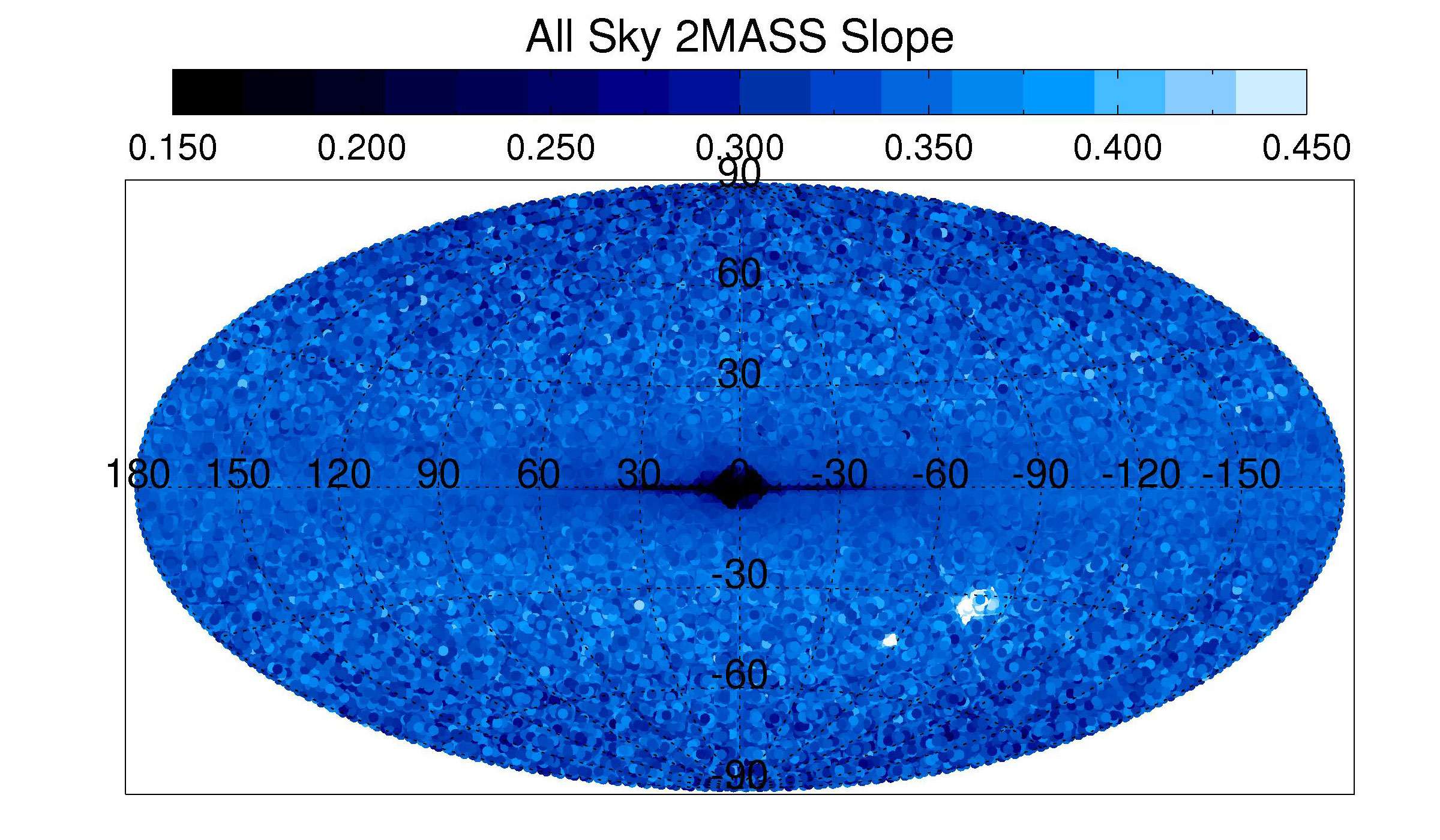}}
 \caption[]{The whole sky 2MASS DSC slopes in magnitude 8 to 14 seem
 homogeneous. The average value is $\sim 0.32$.}
\end{figure}
%----------------------------------------------------------------------

%----------------------------------------------------------------------
\begin{figure} %Fig.~3
% \centerline{{\epsfxsize=4cm\epsffile{thin.eps}\epsfxsize=4cm\epsffile{thick.eps}}}
 \centerline{\includegraphics[width=40mm]{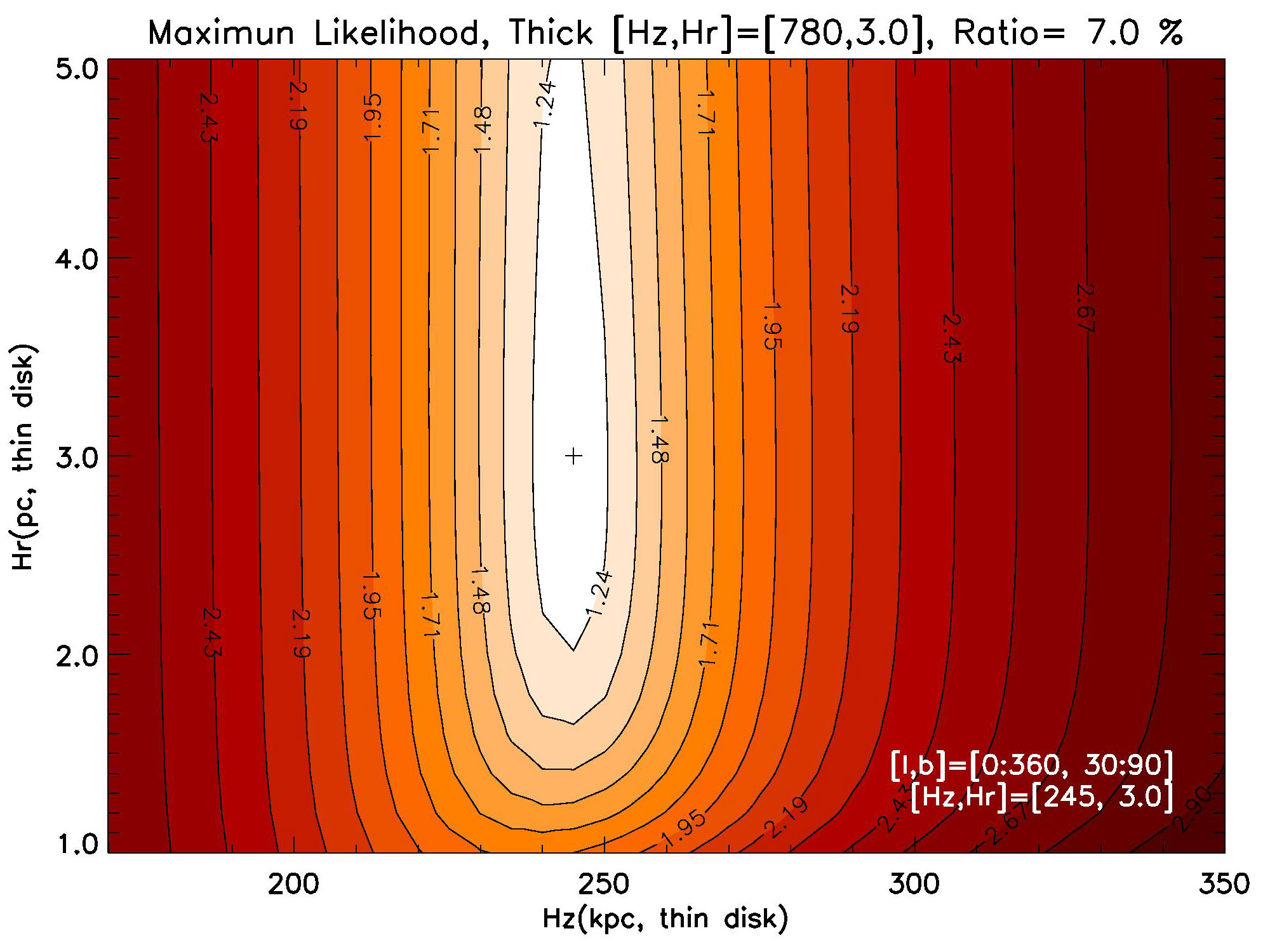}\includegraphics[width=40mm]{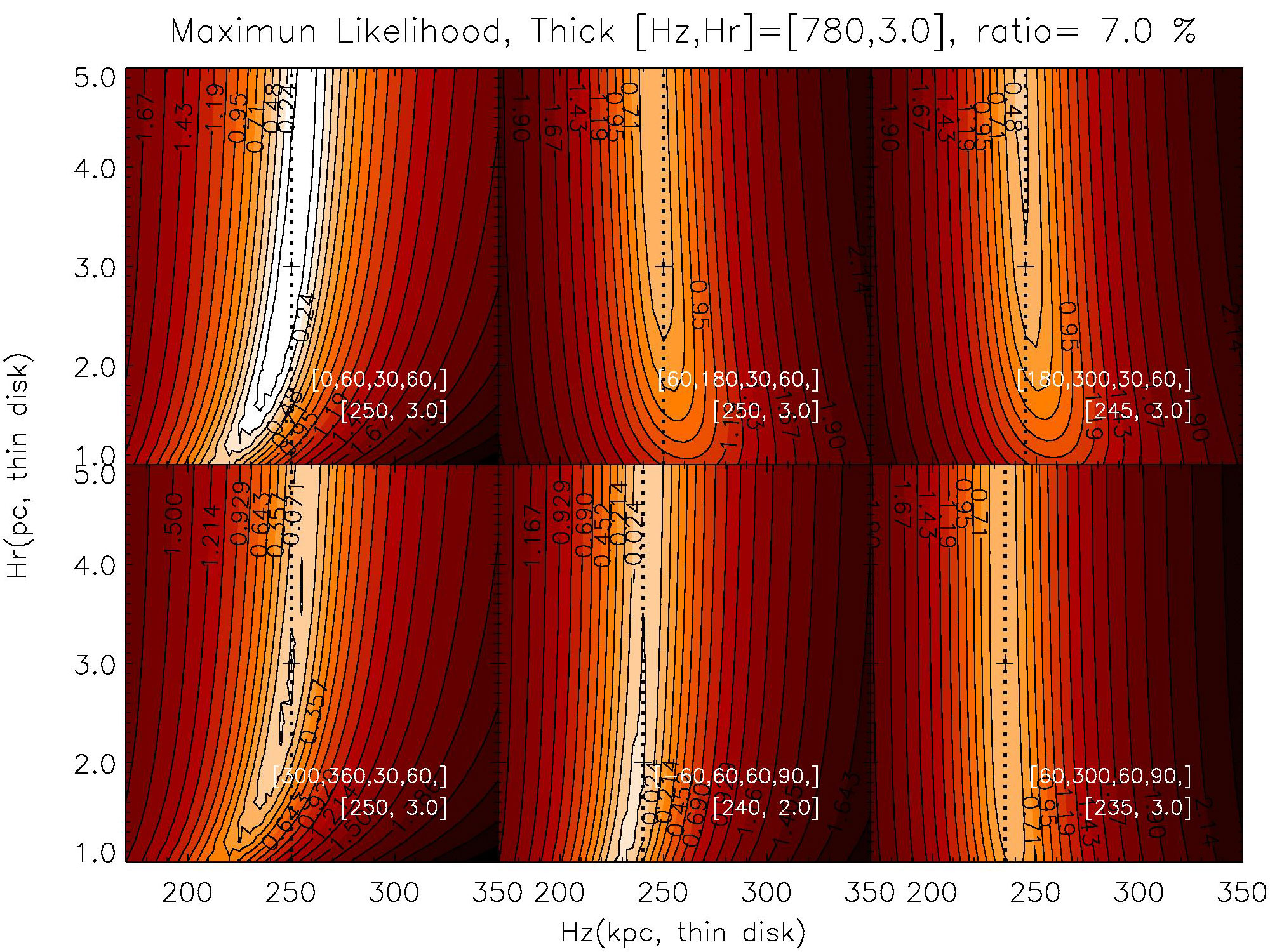}}
 \caption[]{The method of maximum likelihood is used to find $h_r$ and $h_z$.
  The $h_r$ and $h_z$ of the thin disk are $\sim 3.0$ kpc, $\sim 245$ pc
  and that of thick disk are $\sim 3.0$ kpc, $\sim 780$ pc. The ratio of
  the thick to the thin disk is $\sim ~7\%$.}
\end{figure}
%----------------------------------------------------------------------

%----------------------------------------------------------------------
\begin{figure} %Fig.~5
 %\centerline{{\epsfxsize=4cm\epsffile{comparison.eps}\epsfxsize=4cm\epsffile{comparison01.eps}}}
 \centerline{\includegraphics[width=40mm]{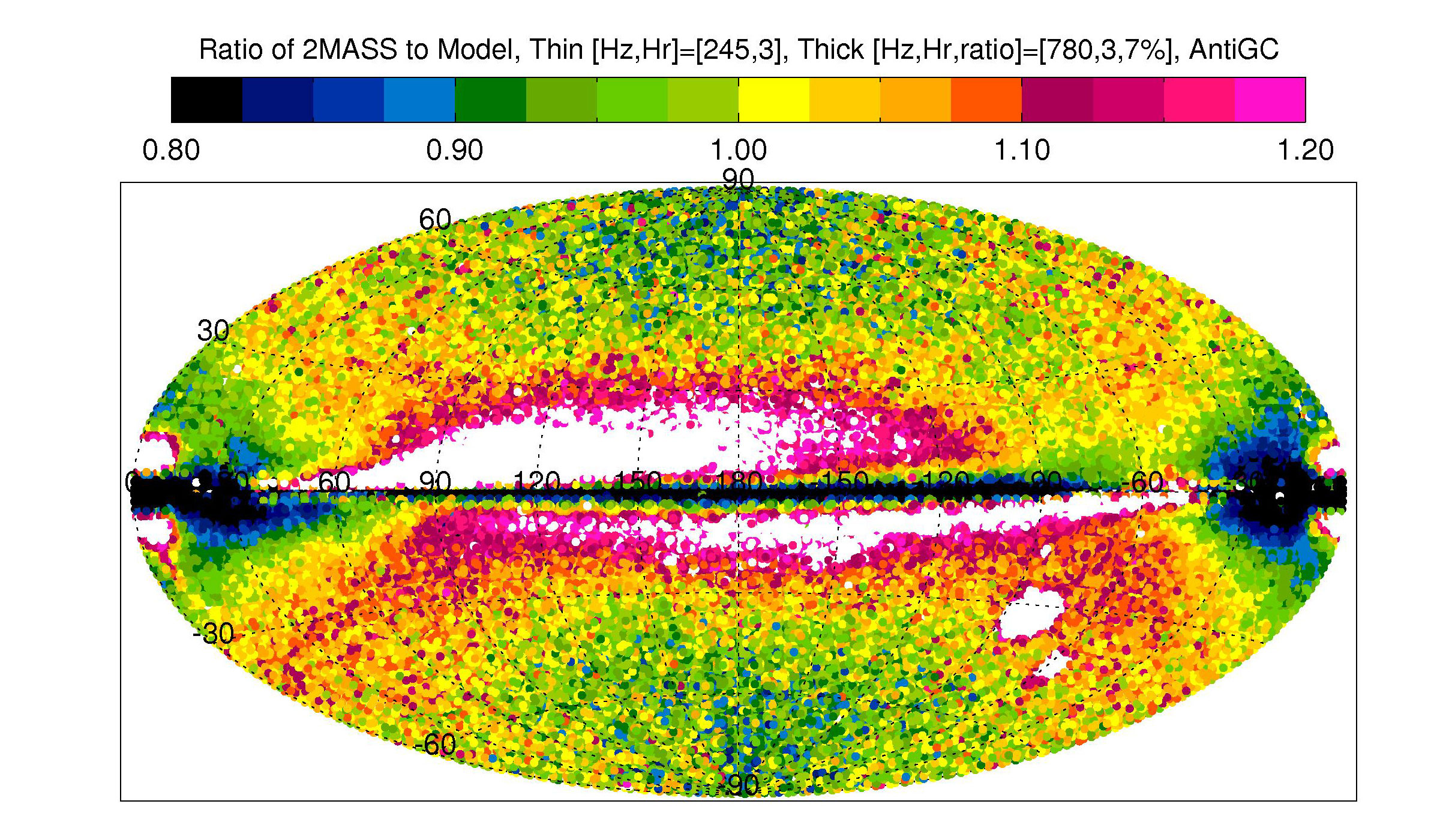}\includegraphics[width=40mm]{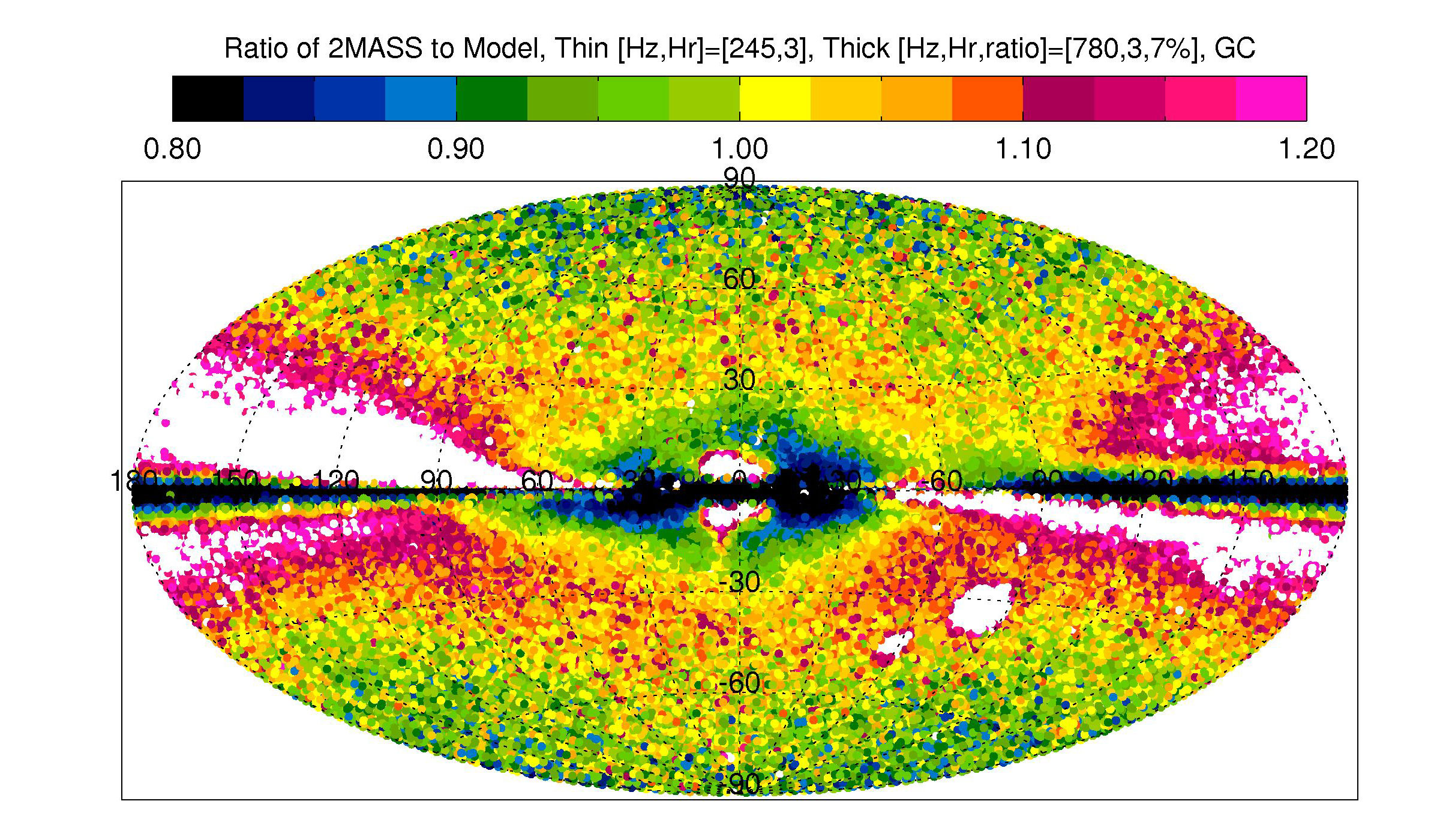}}
 \caption[]{The number ratio of the 2MASS data to the model.
 Left: towards anti-Galactic center. Right: towards the Galactic center.}
\end{figure}
%----------------------------------------------------------------------

\section{Results and Discussion}
\subsection{Luminosity Function}
The differential star counts between $K_s$ magnitude 8 to 14 was
used to derive the slope of the DSC of the whole sky. The slope
seems roughly isotropic over the whole sky, and the average value is
$\sim 0.32$, and hence the power law index of the LF, $\gamma\sim
1.8$ (see Fig.2). If the slope is removed from the DSC, a small but
significant dip appears (see Fig.1 right). Although the depth and
width of the dip are hard to quantify, it can only be reproduced by
some deficiency on the LF. We strongly believed that the dip is not
caused by the fluctuation on density profile. In this case, the
deficiency on LF can help us to quantify $h_r$ and $h_z$, and vice
versa. According to the derived $h_r$ and $h_z$ of the thin disk,
the deficiency on the LF is located at $\sim 0$ Ks absolute
magnitude.

\subsection{Density Profile}
The method of maximum likelihood was applied to find the best fit
$h_r$ and $h_z$. In region with $|b| \geq 30^0$, the thin disk $h_r$
and $h_z$ are $\sim 3.0$ kpc, $\sim 245$ pc, and those of the thick
disk are $~\sim 3.0$ kpc, $\sim 780$ pc. The ratio of the thin to
the thick disk is $\sim 7\%$ (see Fig.3). $Z_\odot$ is derived from
the number ratio of north to south Galaxy, and the value is $\sim
15$ pc.

\subsection{Comparison Between Data and Model}
Fig.4 shows the number ratio of the 2MASS data to the model. Since
we do not include the Galactic bulge in our model, there is an
obvious discrepancy at the Galactic center. Our model generates less
star in the high latitude area. This discrepancy can not be
explained away by adding thick disk in our model. In addition, there
is another discrepancy in the anti-GC direction. It might be a
result of a warp or an increasing $h_z$ with Galactocentric radius
(Cabrera-Lavers et al. 2007). The NCI model has large uncertainty in
low Galactic latitude area, therefore our model produces
unreasonably large star number in these areas. If we believe that
stellar distribution on the disk is smooth, the star count can
constrain the extinction model.

\section*{Acknowledgment}
This work is supported in part by NSC96-2112-M-008-014-MY3. The
authors thank Japanese Virtual Observatory for technical help.

%=========================
%\appendix
%\section[]{}
%\subsection{Subsection title}
%\bsp
%\label{lastpage}
\label{lastpage}
\end{document}